# Optical pump-probe studies of carrier dynamics in few-layer MoS$_2$

Rui Wang, Brian A. Ruzicka, Nardeep Kumar, Matthew Z. Bellus, Hsin-Ying Chiu, and Hui Zhao[a)]
*Department of Physics and Astronomy, The University of Kansas, Lawrence, Kansas 66045, USA*

(Dated: 30 October 2011)

Femtosecond optical pump-probe technique is used to study charge carrier dynamics in few-layer MoS$_2$ samples fabricated by mechanical exfoliation. An ultrafast pump pulse excites carriers and differential reflection of a probe pulse tuned to an excitonic resonance is detected. We find that the spectrum of the differential reflection signal is similar to the derivative of the resonance. The decay time of the signal is in the range of 150 to 400 ps, increasing with the probe wavelength. With a fixed probe wavelength, the signal magnitude is proportional to the pump fluence, while the decay time is independent of it.

Molybdenum disulfide is a transition metal dichalcogenide with an indirect band gap of 1.29 eV.[1] It is composed of monolayers of S-Mo-S that are bound by the weak van der Waals force, while the atoms in each layer are bound strongly by ionic-covalent interactions. The layered structure allows fabrication of atomically-thin films where the quantum confinement can significantly modify the electronic and optical properties. Although few-layer[2–4] and even monolayer[5] MoS$_2$ samples have been fabricated and studied a long time ago, it regained strong interests recently, partially owing to the success of graphene.[6,7]

Very recently, strong photoluminescence (PL) was observed from few-layer MoS$_2$ samples,[8,9] which was attributed to an indirect-to-direct band gap transition that occurs when changing the thickness from bulk to monolayer.[8,9] Such a transition was also confirmed by theoretical calculations[10] and scanning photoelectron microscopy measurements,[11] although a first-principles many-body perturbation theory suggested that the monolayer MoS$_2$ still has an indirect bandgap,[12] but with an unusually large excitonic effect that can account for the observed strong PL. In addition to the possible use in photonic applications, monolayer MoS$_2$ transistors with a $10^8$ on/off ratio and a room-temperature mobility of more than 200 cm$^2$/Vs have been demonstrated,[13] showing great potentials in electronic applications.[14,15] Although samples used in these studies[8,9,13] were prepared by the simple mechanical exfoliation method,[6] techniques with potential for large-scale production, such as liquid exfoliation,[16–18] have been demonstrated to produce high-mobility samples.[19]

For applications in photonics and electronics, it is necessary to understand the dynamics of charge carriers. Previous steady-state optical studies, including PL,[8,9] absorption,[8] reflection,[8,9] photoconductivity,[8] and Raman scattering,[20] have revealed many aspects of electronic and lattice properties. However, time-resolved optical measurements can provide direct information about the carrier dynamics, as illustrated by a recent time-resolved PL measurement.[21] Here we demonstrate that a

---

[a)]Electronic mail: huizhao@ku.edu

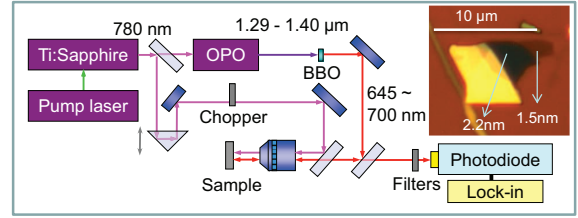

FIG. 1. Schematics of the experimental setup. The inset is a microscopy photo of the studied flake under white light illumination.

femtosecond optical pump-probe technique can be used to study charge carrier dynamics in few-layer MoS$_2$.

Few-layer MoS$_2$ samples are fabricated by mechanical exfoliation with an adhesive tape[6] from natural crystals (SPI Supplies). By using a silicon substrate with a 90-nm SiO$_2$ layer, flakes with different atomic layers can be readily identified by using the contrast of microscopy images.[22,23] The inset of Fig. 1 shows an example of the identified flakes. From atomic-force microscopy measurements, one region has an average thickness of 1.5 nm. Since monolayer MoS$_2$ (one Mo layer sandwiched by two S layers) is 0.65-nm thick, we assign this region to a bilayer, considering uncertainties in the measurement. The region next to it (to the left) has an average thickness of 2.2 nm. It is assigned as a trilayer, which is next to a thick region of many layers (yellowish area).

Figure 1 shows schematically the experimental setup. A diode laser is used to pump a mode-locked Ti:Sapphire laser to generate 100-fs pulses with a central wavelength of 780 nm and a repetition rate of 80 MHz. A small portion of this beam is reflected by a beamsplitter and used as the pump, which is focused to the sample with a spot size of about 2 $\mu$m by using a microscope objective lens. The majority of the 780-nm beam is used to pump an optical parametric oscillator (OPO), which generates a signal output with a central wavelength in the range of 1290 to 1400 nm. This pulse is focused to a beta barium borate (BBO) crystal to generate its second harmonic pulse in the range of 645 to 700 nm, used as the probe pulse. The bandwidth of the probe pulse is about 3.6 nm, corresponding to a pulse width of about 180 fs. The probe pulse is focused to the sample with a spot size of about 4.4 $\mu$m by using the same objective lens. The reflected pulse is collimated and sent to a photodiode, the output of which is detected by a lock-in amplifier. A retroreflector is used in the pump arm in order to control the time delay between the pump and the probe pulses. A mechanical chopper modulates the pump at about 2 KHz for lock-in detection. Reflection and scattering of the pump are prevented from reaching the photodiode by using color filters.

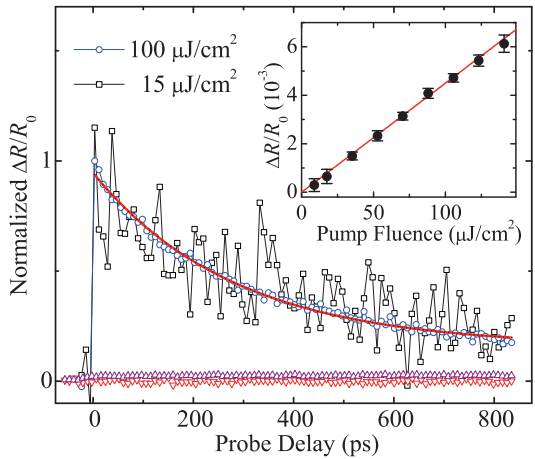

FIG. 2. Normalized differential reflection signals measured with a probe pulse of 660.0 nm and with pump fluences of 100 (circles) and 15 (squares) $\mu J/cm^2$, respectively. The up-triangles and down-triangles are measured from a thick flake and the substrate, respectively. The are normalized against the 100-$\mu J/cm^2$ curve. The inset shows the peak differential reflection signal as a function of the pump fluence.

With this setup, the 780-nm pump pulse injects carriers via phonon-assisted indirect absorption in bilayers[8] or through exciting edge states at the boundary of the flakes. The lock-in amplifier measures the change in the reflection of the probe pulse from the sample induced by these carriers. Such a quantity is defined as differential reflection, $\Delta R/R_0 = (R - R_0)/R_0$, where $R$ and $R_0$ are the reflection of the probe pulse from sample with and without the presence of the pump pulse, respectively. All the measurements are carried out under ambient conditions and at room temperature.

First, we use a probe pulse with a central wavelength of 660 nm. Such a wavelength is in the short-wavelength side of the excitonic resonance of the sample (centered at 672.6 nm, see the PL spectrum in Fig. 3B). The differential reflection signal is measured as we vary the energy fluence of the pump pulse. The blue circles in Fig 2 show the normalized signal measured with a pump fluence of about 100 $\mu J/cm^2$. It raises to the peak right after zero delay, then decays exponentially, with a time constant of 281 ps. The normalized signal measured with an 15-$\mu J/cm^2$ pump pulse (black squares) has the same shape, indicating the decay time is independent of the carrier density. The inset of Fig. 2 shows that the peak signal, on the order of $10^{-3}$, is proportional to the pump fluence. Such a fluence dependence confirms that the 780-nm pump pulse excites carriers via one-photon absorption, instead of two-photon absorption, since the latter would give a quadratic fluence dependence. The positive signal means the reflection of the probe pulse is enhanced by the carriers, which indicates the absorption is reduced. In these measurements, the laser spots are located on the bilayer and trilayer regions. A much weaker signal was detected when they are moved to the central part of thick flakes, as indicated by the up-triangles in Fig. 2. Since the excitonic resonance is similar in bulk and few-layer flakes,[8,9] this suggests that the 780-nm pulse does not efficiently inject carriers in the bulk. Furthermore, no signal was detected when the laser spots are moved to the parts of the substrate that are not covered by $MoS_2$ (down-triangles in Fig. 2).

Next, we tune the probe wavelength to study how the detuning from the excitonic resonance influences the differential reflection signal. Figure 3A shows the measured signal as a function of the probe delay for many probe wavelengths, which are all in the excitonic resonance, as confirmed by the PL spectrum shown in Fig. 3B. When the probe wavelength is increased (from bottom to top in Fig. 3A), the magnitude of the signal changes dramatically and the sign changes from positive to negative. This trend is summarized in Fig. 3C (squares), where the peak signal magnitude near zero delay is plotted as a function of the probe wavelength. The shape is similar to the derivative of the Gaussian-shaped excitonic PL spectrum.

The strong dependence of the signal on the detuning between the probe and the excitonic resonance shown in Figs. 3A and 3C provides unambiguous evidence that the differential reflection signal is caused by changes in the excitonic resonance. The derivativelike shape indicates that the dominant change is a shift of the central wavelength. We attempt to fit the spectrum by taking the difference between the Gaussian spectrum (red line in Fig. 3B) and an identical one but slightly shifted by an adjustable amount. The best fit is plotted as the blue dashed line in Fig. 3C, with a shift towards long wavelength of 0.162 nm. The discrepancy between the fit and the data suggests that the shift is not the only change on the resonance. By also including changes in the height and the width of the resonance, we are able to fit the data: The red solid line in Fig. 3C is obtained by subtracting the original Gaussian function by another one that is 0.37($\pm$0.15)% higher, 0.57($\pm$0.15)% narrower, and with a central wavelength 0.14($\pm$0.03)-nm longer. Due to the large uncertainties on the data and the limited spectral range constrained by the laser tuning range, we do not view these values as accurate determinations of these three contributions. Rather, these numbers suggest that with reasonable values, the spectrum can be described by the changes to the excitonic resonance.

The decay time of the signal also changes dramatically with the probe wavelength, as summarized in Fig. 3D. Despite the relatively low signal level and the limited scan range of the probe delay, the trend of increase is conclusive, as clearly illustrated by the two curves shown in Fig. 3E. We attribute this to different dynamics of the three contributions, since different wavelengths probe different combinations of the three contributions.

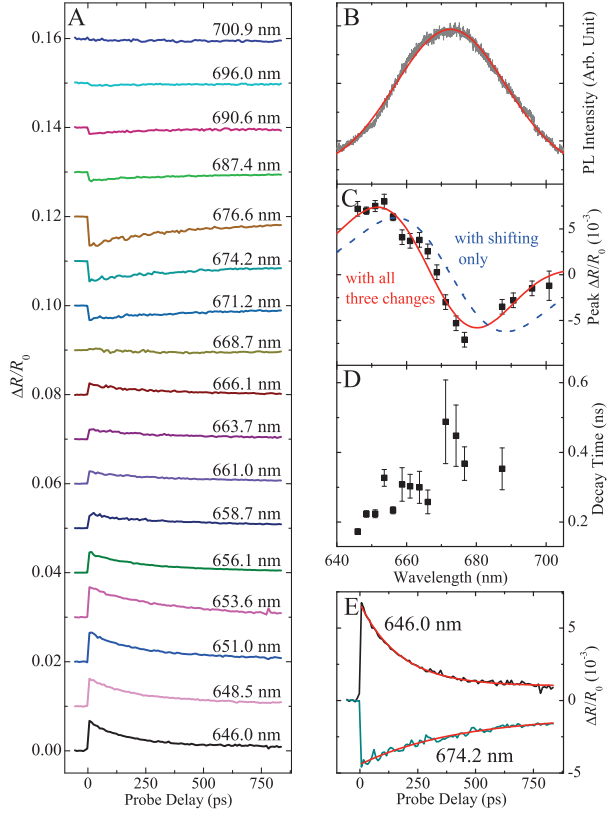

FIG. 3. A: Differential reflection signals measured with different probe wavelengths. The curves are shifted for clarity. B: PL spectrum measured with a excitation of 404 nm. The red curve shows a Gaussian fit. C: Peak magnitude of differential reflection signal as a function of the probe wavelength. The red solid line is a fit including changes in height, width, and central wavelength of the Gaussian function, while the blue dashed line includes only the shift. D: The decay time of the differential reflection signal as a function of the probe wavelength, deduced from single-exponential fits to the curves shown in A. E: Two of the curves shown in A are re-plotted with the fits.

In summary, we demonstrate that a femtosecond pump-probe technique can be used to study charge carrier dynamics in few-layer $MoS_2$ samples. Carriers injected by a 780-nm pump pulse can be efficiently detected by measuring the differential reflection of a probe pulse tuned to the excitonic transition near 670 nm. We found that the magnitude, the sign, and the the decay time of the signal change dramatically as the probe wavelength is tuned within the excitonic resonance. Besides providing quantitative information on the carrier dynamics in this important new two-dimensional material, our experiment may stimulate further optical studies of carrier dynamics in this material system.

We would like to thank Chih-Wei Lai and Shenqiang Ren for many helpful discussions, and Rodolfo Torres-Gavosto and Cindy Berry for their help on AFM measurements. We acknowledge support from the US National Science Foundation under Awards No. DMR-0954486 and No. EPS-0903806, and matching support from the State of Kansas through Kansas Technology Enterprise Corporation.